\documentclass[11pt]{article}
\pdfoutput=1
\usepackage{jheppub} 

\usepackage[T1]{fontenc} 
\usepackage[utf8]{inputenc}
\usepackage[normalem]{ulem}

\usepackage{tikz}
\usepackage{pifont}
\usepackage{enumerate}
\usepackage[shortlabels]{enumitem}
\usepackage{amsfonts}
\usepackage{amssymb}
\usepackage{amsmath}
\usepackage{amsthm}
\usepackage{latexsym}

\usepackage{tensor}
\usepackage{xcolor}

\newcommand{\be}{\begin{equation}}
\newcommand{\ee}{\end{equation}}
\newcommand{\ba}{\begin{eqnarray}}
\newcommand{\ea}{\end{eqnarray}}

\newcommand{\rh}{r_{+}}
\par
\title{\textbf{Thermodynamics of Rotating NUTty Dyons}}

\author[a,b]{Alvaro Ballon Bordo}

\author[a,b]{Finnian Gray}

\author[a,b]{David Kubiz\v n\'ak}

\affiliation[a]{Perimeter Institute, 31 Caroline Street North, Waterloo, ON, N2L 2Y5, Canada}
\affiliation[b]{Department of Physics and Astronomy, University of Waterloo,
	Waterloo, Ontario, Canada, N2L 3G1}

\emailAdd{aballonbordo@perimeterinstitute.ca}
\emailAdd{fgray@perimeterinstitute.ca}
\emailAdd{dkubiznak@perimeterinstitute.ca}
\abstract{
We extend the work on thermodynamics of Lorentzian NUTty solutions, by including simultaneously the effect of rotation and electromagnetic charges.
Due to the fact that Misner strings carry electric charge, and similar to the non-rotating case, we observe an interesting interplay between the horizon and asymptotic charges. Namely, upon employing the Euclidean action calculation we derive two alternative full cohomogeneity first laws, one that includes the variations of the horizon electric charge and the asymptotic magnetic charge and another that involves the horizon magnetic charge and the asymptotic electric charge. When one of the horizon charges vanishes, we obtain the corresponding `electric' and `magnetic' first laws that are connected by the electromagnetic duality. We also briefly study the free energy of the corresponding system.

}
\begin{document}
\maketitle

\section{Introduction}

The Lorentzian Taub--NUT spacetime \cite{taub1951empty, newman1963empty} is considered to be an exotic solution to Einstein's equations, mainly due to the presence of a hitherto unobserved gravitomagnetic charge. This so-called NUT-charge endows the spacetime with rotating {\em string-like singularities} (Misner strings) \cite{bonnor1969new, manko2005physical}
that are accompanied by regions with closed timelike curves in their vicinity, raising serious doubts on the physical plausibility of the solution.
These issues are non-existent if we periodically identify the time coordinate \cite{misner1963flatter}, and for many years the study of Taub--NUTs and their  thermodynamics was restricted to the Euclidean solutions, see for example \cite{Page:1979aj, Page:1979zv, Hawking:1998jf, Hawking:1998ct, Chamblin:1998pz, Emparan:1999pm, Mann:1999pc, Mann:1999bt, Johnson:2014xza, Johnson:2014pwa, Garfinkle:2000ms}. Nevertheless, in the light of some recent arguments that show that the Lorentzian solutions {(without imposing the time periodiciy condition)} are actually well-behaved for geodesic curves \cite{Clement:2015cxa, Clement:2015aka, Clement:2016mll} {(see also \cite{miller1971taub} for the Kruskal extension over the black hole horizon)} the interest in the Lorentzian thermodynamics has currently gained momentum \cite{Kubiznak:2019yiu, Durka:2019ajz, Ballon:2019uha, Bordo:2019tyh, Bordo:2019rhu, Clement:2019ghi, Wu:2019pzr, Chen:2019uhp}.

Namely, a full cohomogeneity first law has been recently proposed for the Lorentzian Taub-NUT solution in the asymptotically Anti-de Sitter case \cite{Kubiznak:2019yiu}, and for its (charged) dyonic generalization \cite{Ballon:2019uha}, as well as for the rotating asymptotically flat Taub-NUT spacetimes \cite{Bordo:2019rhu}. The key ingredient to formulate these first laws is to abandon the time periodicity condition and to identify the entropy of the system with one quarter of the black hole horizon area. For this to work, one needs to  include variations of the Misner charge \cite{Kubiznak:2019yiu, Bordo:2019tyh}, a quantity conjugate to the {\em Misner potential}, {where the latter is either} identified with the surface gravity of the string \cite{Bordo:2019tyh} or equivalently with its angular velocity \cite{Durka:2019ajz, Clement:2019ghi}. In the case of dyonic solutions, one also observes an interesting interplay between the horizon and asymptotic magnetic and electric charges, related to the fact that Misner strings carry electric charge \cite{Ballon:2019uha}.

The purpose of this work is to extend the above results and formulate the first law for the rotating NUTty dyons in the asymptotically flat case. In order to do this, we employ the Euclidean action method and propose new expressions for the Misner charge and the angular momentum, calculated as conjugate variables to the Misner potential and the horizon angular  velocity, respectively. Similar to the non-rotating case, we still observe the mixing of electric and magnetic charges at the horizon and at infinity. Upon requiring that the due horizon charge vanishes, we write down the purely electric and magnetic laws for the rotating case  and study the behavior of the resultant free energy.

Our work is organized as follows. In Sec.~2, we  review the rotating NUTty solution with Lorentzian signature and calculate some of the relevant thermodynamic quantities. Sec.~3 is devoted to calculating the conjugate thermodynamic variables from the free energy for the full dyon. As a result, we obtain two versions of a full cohomogeneity first law both of which mix the horizon and asymptotic charges. In Sec.~4, we simplify the general case to write down purely electric and magnetic laws. Sec.~5 is devoted to conclusions. We refer the interested reader to App.~A, where we lay out the details of the calculation of the free energy for the rotating NUTty dyon.

\section{Rotating NUTty dyons}
The geometry of a rotating NUTty dyon can be obtained by a certain limit of the general type D Plebanski--Demianski spacetime \cite{plebanski1976rotating}. It reads
\begin{equation}\label{eq:Met}
ds^2=-\frac{\Delta}{U} (e^1)^2+\frac{U}{\Delta}dr^2+Ud\theta^2+\frac{\sin^2\!\theta}{U} (e^2)^2\;,
\end{equation}
where $e^1$ and $e^2$ are the following basis 1-forms:
\begin{equation}
e^1=dt+\bigl[2 (s +n\cos\theta)-a \sin^2\!\theta\bigr]d\varphi\,,\quad
e^2=a dt-\left(r^2+n^2+a^2-2 a s \right)d\varphi\,,
\end{equation}
and the metric functions are given by
\begin{align}
	\Delta&=a^2+e^2+4 g^2 n^2-n^2-2 m r+r^2\;,\\
U&=r^2+(n+a\cos\theta)^2\,.
	\end{align}
Here, the parameters $a, e, g, n, m$ are the rotation, electric charge, magnetic charge, NUT, and mass parameters. The parameter $s$ governs the distribution of Misner strings, e.g. \cite{Clement:2015cxa, Bordo:2019tyh}. In particular, when
 \be
 s=0\,,
\ee
the strings  are symmetrically distributed. Since in this case many of the thermodynamic formulae simplify, we shall commit to this choice for the rest of this paper.

The gauge potential is given by
\begin{equation}\label{eq:A}
A=-\frac{e r+g(r^2+a^2-n^2)}{U}e^1+\frac{ag\sin^2\!\theta}{U}e^2\;.
\end{equation}
and the corresponding field strength is $F=dA$. With this, we have a solution of electrovacuum Einstein equations (setting $G=1$),
\begin{align}
R_{ab}-\frac{1}{2}R g_{ab}&={8\pi}T_{ab}\;,\quad T_{ab}={\frac{1}{4\pi}}\left(F_{ac}F^{\;\;\;c}_{b}-\frac{1}{4}g_{ab}F_{cd} F^{cd}\right)\,,\\
d*F&=0\;.
\end{align}
Note that, upon using the electromagnetic duality:
\be\label{EMduality}
\mbox{electric}\leftrightarrow\mbox{magnetic}:\quad
e\leftrightarrow-2ng\,,\quad 2ng\leftrightarrow e\,,
\ee
we can alternatively write the ``dual'' gauge potential
\begin{equation}\label{eq:B}
B=-\frac{-4n^2gr+e(r^2+a^2-n^2)}{2nU}e^1+\frac{ae\sin^2\!\theta}{2nU}e^2\,,
\end{equation}
which yields $*F=-dB$\,.

The spacetime is stationary and axisymmetric, the  corresponding timelike and angular Killing vectors being $\partial_t$ and $\partial_\varphi$.
The conformal (asymptotic) mass and angular momentum associated with these Killing vectors are given by
\begin{equation}
M=m=\frac{a^2+e^2+4 g^2n^2-n^2+\rh^2}{2 \rh}\,,\quad J_{\infty} =Ma\,.
\end{equation}

The spacetime admits three horizons; one associated to the {\em black hole} located at the largest root of $\Delta(r_+)=0$, and generated by the Killing vector
\begin{equation}\label{eq:BHxi}
\xi_H=\partial_t+\Omega_H \partial_\varphi\,.
\end{equation}
It is to this horizon that we associate the black hole temperature and entropy
\begin{align}
	T&=\frac{\Delta'(r_+)}{4\pi(r^2_++n^2+a^2)}=\frac{1}{4 \pi \rh}\frac{r_+^2+n^2-a^2-e^2-4 g^2 n^2}{a^2+n^2+\rh^2}\,,\label{T}\\
	S&=\frac{\text{Area}}{4}=\pi(r_+^2+n^2+a^2)\,.\label{S}
\end{align}
The other two are located along the north and south pole axis $\cos\theta=\pm1$, i.e., along the Misner strings and generated by the Killing vectors
\begin{equation}\label{eq:Tubexi}
\xi_\pm=\partial_t+\Omega_\pm \partial_\varphi\,.
\end{equation}
Here, the angular velocities are found by evaluating the angular velocity of the ``Zero Angular Momentum Observers" (ZAMOs)
\begin{equation}
	\omega=-\frac{g_{t\varphi}}{g_{\varphi\varphi}}
\end{equation}
on each horizon
\begin{align}
	\Omega_H&=\omega\bigg|_{r=r_+}=\frac{a}{r^2_++n^2+a^2}\,,\\
	\Omega_\pm&=\omega\bigg|_{y=\pm1}=\mp\frac{1}{2n}\;,
\end{align}
while the angular velocity at infinity, $\Omega_\infty=\omega|_{r=\infty}$, vanishes.
Note that we can also formally associate surface gravities to the Misner string horizons, given by
\be
\kappa_\pm=\sqrt{\left|\frac{1}{2}(\nabla_\alpha \xi^\pm_\beta) (\nabla^\alpha \xi_\pm^\beta)\right|}=\frac{1}{2 n}\,.
\ee
Up to a constant factor, these coincide with the angular velocities $\Omega_\pm$. This dual meaning begs the question of whether we should interpret the Misner potential
\be
\psi=\frac{1}{8\pi n}
\ee
as a string temperature, $\kappa_\pm/(4\pi)$, or as angular velocity, $|\Omega_\pm|/(4\pi)$. The latter perspective is argued for in \cite{Clement:2019ghi} in the context of rotating Taub-NUT thermodynamics, where the Misner charges are interpreted as the string angular momenta (see also \cite{Durka:2019ajz}).

Similar to the non-rotating case, we find that Misner strings carry charges and the field strength integral over a 2-sphere depends on the radius. Namely, we find
\begin{align}
	Q_e(r)&=\frac{1}{4\pi}\int_{S^2} *F=\frac{\left(a^2 +n^2+r^2\right) \left[e \left(r^2+a^2-n^2\right)-4 g n^2 r\right]}{\left[(a-n)^2+r^2\right] \left[(a+n)^2+r^2\right]}\;,\\
Q_m(r)&=\frac{1}{4\pi}\int_{S^2} F=-\frac{2 n\left(a^2 +n^2+r^2\right) \left[g \left(a^2-n^2+r^2\right)+e r\right]}{\left[(a-n)^2+r^2\right] \left[(a+n)^2+r^2\right]}\;.
\end{align}
For any given radius $r$, these are related by the electromagnetic duality \eqref{EMduality}.
Specifically, we can define the horizon electric and magnetic charges
\begin{align}
	Q_e^H&=Q_e(r=r_+)=\frac{\left(a^2 +n^2+r_+^2\right) \left[e \left(r_+^2+a^2-n^2\right)-4 g n^2 r_+\right]}{ \left[(a-n)^2+r_+^2\right] \left[(a+n)^2+r_+^2\right]}\;,\\
Q_m^H&=Q_m(r=r_+)=-\frac{2 n\left(a^2 +n^2+r_+^2\right) \left[g \left(a^2-n^2+r_+^2\right)+e r_+\right]}{\left[(a-n)^2+r_+^2\right] \left[(a+n)^2+r_+^2\right]}\;,
\end{align}
and their asymptotic analogues
\begin{align}
Q_e&=\lim_{r\to \infty}Q_e(r)=e\;,\label{eq:Qeinfty}\\
Q_m&=\lim_{r\to \infty} Q_m(r)=-{2ng}\;.
\end{align}

To calculate the thermodynamic electrostatic potential $\phi_e$ we will employ the  Hawking--Ross prescription \cite{Hawking:1995ap}. The first step is to find a gauge, by applying
$A\to A+\Phi^e_t dt$, to ensure that the electrostatic potential defined by  $\Phi_e(r)=-\xi^a A_a$ vanishes on the horizon, $\Phi_e(r_+)=0$. To achieve this we find\footnote{A gauge transformation $A\to A+ \Phi_\varphi d\varphi$ changes distribution of Dirac strings. One can easily check that with our choice of gauge, the integral $\int_{S^2_{\infty}} A_\phi *F$ vanishes. This means that the Dirac strings are symmetrically distributed and do not induce any asymptotic Komar angular momentum \cite{Clement:2019ghi}.}
\be
\Phi^e_t=\frac{g(a^2-n^2+r_+^2)+er_+}{a^2 +n^2+r_+^2}\,.
\ee
The next step is to evaluate the (Euclidean) Hawking--Ross boundary integral, which yields the thermodynamic potential
\begin{align}
\phi_e=\frac{1}{4\pi \beta Q}\int_{\partial M} \sqrt{-h}\;n_\mu F^{\mu\nu}{A}_{\nu}\,.
\end{align}
Here, $\partial M$ is a hypersurface of $r=const$ at infinity, $\beta=1/T$, $Q$ is identified with the asymptotic electric charge $Q_e$, and we have to use the above gauge. This yields
\be
\phi_e=\frac{er_+-2gn^2}{a^2+n^2+r_+^2}\,,
\ee
which straightforwardly generalizes the electrostatic potentials discussed in \cite{Ballon:2019uha} for the non-rotating case. Alternatively, this potential can be found with the usual prescription
\be
 \phi_e=-\left(\left.\xi^a A_a\right|_{r=\rh}-\left.\xi^a A_a\right|_{r=\infty}\right),
\ee
evaluated at $\theta=\pi/2$.

\section{Full cohomogeneity first laws}

\subsection{Possibility 1: horizon magnetic charge}
In order to write down the first law of thermodynamics for rotating NUTty dyons, we have employed the technique of Euclidean action calculation. Although the Euclidean solution is `plagued' with orbifold singularities \cite{Clement:2019ghi}, the action is finite. Calculated in Appendix~\ref{AppFE}, it yields the following free energy:
\begin{equation}\label{G}
{G}= \frac{m}{2}-\frac{\rh}{2}\frac{(\rh^2-n^2+a^2) \left(e^2-4 g^2 n^2\right)-8 e g n^2 \rh}{\left[\rh^2+(a+n)^2\right]\left[\rh^2+(a-n)^2\right]}\,.
\end{equation}

As with the non-rotating dyon we have now two possibilities: i) treat the calculated free energy $G$ as a function of the magnetic horizon charge $Q_m^H$, or ii) treat it as a function of the asymptotic magnetic charge $Q_m$. The first choice is
\begin{equation}\label{G1}
G=G(T,\psi, \Omega_H,\phi_e,Q_m^H)\,,
\end{equation}
and defines the following thermodynamic quantities:
\begin{align}
S=-\frac{\partial { G}}{\partial T}\;,\quad N=-\frac{\partial {G}}{\partial \psi}\;,\quad
J=-\frac{\partial {G}}{\partial \Omega_H}\;,\quad
Q_e= -\frac{\partial {G}}{\partial \phi_e}\;,\quad \phi_{m}=\frac{\partial {G}}{\partial Q_m^H}\;.
\end{align}
This yields the correct entropy $S$, \eqref{S}, and the asymptotic electric charge, $Q_e$. The magnetic potential is
\be\label{phim}
\phi_{m}=\hat \phi_m+2n \Omega_H^2 Q_e\,,
\ee
where we have denoted by $\hat \phi_m$ the electromagnetic dual to the electrostatic potential $\phi$,
\be
\hat \phi_m=-\frac{n(2gr_++e)}{a^2+n^2+r_+^2}\,.
\ee
Observe that the magnetic potential $\phi_m$ is in the presence of rotation no longer a dual to $\phi$.
The Misner charge $N$ and the conjugate angular momentum $J$ are given by
\begin{align}
N =&2 n\pi\left[\frac{4 e \left(e \rh-2 g n^2\right)}{a^2+n^2+\rh^2}-\frac{2 n^2 \left(a^2+e^2+\left(4 g^2-1\right) n^2\right)}{\rh (a-n) (a+n)}\right.\nonumber\\
& +\bigg(\frac{2 n (a+n) \left(4 e g n (a+n)-\rh \left(e^2-4 g^2 n^2\right)\right)}{\left((a+n)^2+\rh^2\right)^2}+(n\leftrightarrow -n)\bigg)\nonumber\\
&+\left.\bigg(\frac{4 a e g n (a+n)-\rh \left(2 a e^2+3 e^2 n+4 g^2 n^3\right)}{\rh^2 (a+n)+(a+n)^3}+(n\leftrightarrow -n)\bigg)\right)\,,\\
J =&\frac{1}{2}\left[\frac{a \left(a^2+n^2\right) \left(a^2+e^2+\left(4 g^2-1\right) n^2\right)}{\rh (a-n) (a+n)}+a \rh \right.\nonumber\\
&+\bigg(\frac{2 n^2 (a+n) \left(4 e g n (a+n)-\rh \left(e^2-4 g^2 n^2\right)\right)}{\left((a+n)^2+\rh^2\right)^2}+(n\leftrightarrow -n)\bigg)\nonumber\\
&+\left.\bigg(\frac{n \left(4 a e g n (a+n)-\rh \left(2 a e^2+3 e^2 n+4 g^2 n^3\right)\right)}{\rh^2 (a+n)+(a+n)^3}+(n\leftrightarrow -n)\bigg)\right]\,.
\end{align}
The derived quantities satisfy the following first law and Smarr relations:
\begin{align}
\delta M&=T\delta S+\Omega_H \delta J+ \phi_e \delta Q_e+\phi_m \delta Q_m^H+\psi\delta N\,,\\
M&=2(TS+\psi N+\Omega_H J)+\phi_e Q_e+\phi_m Q_m^H\,.\label{SmarrFull1}
\end{align}
One can easily check that
\be
G=\frac{1}{2}\Bigl(M+\phi_m Q_m^H-\phi_e Q_e\Bigr)\,,
\ee
which, upon using the Smarr relation \eqref{SmarrFull1}, is equivalent to
\be
G=M-TS-\phi_e Q_e-\psi N-\Omega_H J\,.
\ee


\subsection{Possibility 2: asymptotic magnetic charge}
The second possibility is to associate ${G}$ , \eqref{G}, with
\begin{equation}\label{G2}
\tilde {G}=\tilde {G}(T,\psi, \Omega_H,\phi_e, Q_m)\,,
\end{equation}
and defines the following thermodynamic quantities:
\begin{align}
S=-\frac{\partial \tilde {G}}{\partial T}\;,\quad \tilde N=-\frac{\partial \tilde {G}}{\partial \psi}\;,\quad
\tilde J=-\frac{\partial \tilde {G}}{\partial \Omega_H}\;,\quad
\tilde Q_e= -\frac{\partial \tilde { G}}{\partial \phi_e}\;,\quad \tilde \phi_{m}=\frac{\partial \tilde {G}}{\partial Q_m}\;.
\end{align}
This again yields the correct entropy $S$, \eqref{S}, but now leads to the horizon electric charge, $\tilde Q_e=Q_e^H$, and a new
magnetostatic potential
\be
\tilde \phi_m=\hat \phi_m+2n \Omega_H^2 Q_e^H\,.
\ee
The new $\tilde N$  and $\tilde J$ are given by
\begin{align}
\tilde N =& -2\pi n^2\left[\bigg(\frac{2 (a+n) \left(4 e g n (a+n)-\rh \left(e^2-4 g^2 n^2\right)\right)}{\left((a+n)^2+\rh^2\right)^2}-(n\leftrightarrow -n)\bigg)\right.\nonumber\\
&+\bigg(\frac{\rh \left(e^2+4 g^2 n^2\right)-4 e g n (a+n)}{\rh^2 (a+n)+(a+n)^3}-(n\leftrightarrow -n)\bigg)
 + \left. \frac{2 n \left(a^2+e^2+\left(4 g^2-1\right) n^2\right)}{\rh \left(a^2-n^2\right)}\right]\,,\\
\tilde J =&\frac{1}{2}\left[\bigg(\frac{2 n^2 (a-n) \left(4 e g n (a-n)+\rh \left(e^2-4 g^2 n^2\right)\right)}{\left((a-n)^2+\rh^2\right)^2}+(n\leftrightarrow -n)\bigg)\right.\nonumber\\
&+\bigg(\frac{n^2 \left(4 e g (a+n) (a+2 n)+4 g^2 n \rh (2 a+n)-e^2\rh\right)}{\rh^2 (a+n)+(a+n)^3}+(n\leftrightarrow -n)\bigg)\nonumber\\
&\left.+\frac{a \left(a^2+n^2\right) \left(a^2+e^2+4 g^2n^2-n^2\right)}{\rh (a-n) (a+n)}+a \rh \right]\,.
\end{align}
We have
\be
\tilde J=J-8n^2\Omega_H\tilde \phi_m Q_m^H\,.
\ee
Note also that $\{N, \tilde N\}$ and $\{J, \tilde J\}$ are not electromagnetically dual. Namely, denoting by $\hat N$ the dual to $N$ and similarly for $J$, we find
\be
\tilde N=\hat N-32\pi g n^3 \Omega_H^2Q_e^H\,,\quad \tilde J=\hat J+8n^2g\Omega_H Q_e^H\,.
\ee
The new quantities satisfy the following first law and Smarr relations:
\begin{align}
\delta M&=T\delta S+\phi_e \delta Q_e^H+\Omega_H \delta \tilde J+\tilde \phi_m \delta Q_m+\psi\delta \tilde N\,,\label{Full2}\\
M&=2(TS+\psi \tilde N+\Omega_H \tilde J)+\phi_e Q_e^H+ \tilde \phi_m Q_m\,.
\end{align}
One can easily check that now
\be
\tilde G=\frac{1}{2}\bigl(M+\tilde \phi_m Q_m-\phi_e Q_e^H\bigr)\,,
\ee
which is equivalent to
\be
\tilde {G}=M-TS-\phi_e Q_e^H-\psi \tilde N-\Omega_H \tilde J\,.
\ee

Let us finally mention that an alternative (horizon electric) Smarr relation has recently been derived in \cite{Clement:2019ghi} by employing the Komar integration method, see formula (3.60) therein. Instead of asymptotic magnetic charge, it features Misner string charges $Q_\pm$ and associated electric potentials $\Phi_\pm$. Interestingly, we have checked that the therein presented quantities satisfy the following differential relation:
\be\label{alter}
\delta M=T\delta S+\phi_e \delta Q_e^H+\Omega_H \delta J_H+\Omega_+\delta \tilde J_++\Omega_- \delta \tilde J_-+\Phi_+\delta Q_++\Phi_-\delta Q_-\,,
\ee
where $J_H$ is different from $\tilde J$ presented in our paper. Since $\Omega_+=-\Omega_-$ and $\Phi_+=\Phi_-=g$, the last four terms can be combined into two, and we recover
\be
\delta M=T\delta S+\Omega_H \delta J_H+\phi_e \delta Q_e^H+\Omega_+\delta (\tilde J_+-\tilde J_-)+\Phi_+\delta (Q_++Q_-)\,,
\ee
which represents a full cohomogeneity alternative to our first law \eqref{Full2}. It would be interesting to see whether such thermodynamics can also be derived starting from some Euclidean action.

\section{Electric and magnetic first laws}
Let us now restrict ourselves to two important physical subcases and find a generalization of the electric and magnetic laws presented in \cite{Ballon:2019uha}. The key idea is to constrain the above non-degenerate first laws by imposing a requirement that the corresponding horizon charge vanishes. Namely, the electric first law is obtained by considering the first possibility above and setting the horizon magnetic charge to zero,
\be\label{1a}
Q_m^H=0\quad \Rightarrow \quad g=-\frac{er_+}{r_+^2+a^2-n^2}\,.
\ee
Upon doing this, the magnetic piece in \eqref{G1} vanishes and the following {\em electric-type relations} hold:
\begin{align}
\delta M&=T\delta S+\Omega_H \delta J+ \phi_e \delta Q_e+\psi\delta N\,,\\
M&=2(TS+\psi N+\Omega_H J)+\phi_e Q_e\,.
\end{align}
The thermodynamic quantities simplify significantly and we have
\begin{align}
G&=\frac{m}{2}-\frac{r_+ e^2}{2(r_+^2+a^2-n^2)}=\frac{1}{2}(m-\phi_e Q_e)\,,\label{Gel}\\
M&=m=\frac{r_+^2+a^2-n^2}{2r_+}+\frac{e^2[(r_+^2+a^2+n^2)^2-4a^2n^2]}{2r_+(r_+^2+a^2-n^2)^2}\,.\\
\phi_e&=\frac{er_+}{r_+^2+a^2-n^2}=-g\,,\quad Q_e=e\,,\\
S&=\pi (r_+^2+a^2+n^2)\,,\quad T=\frac{r_+^2+n^2-a^2-e^2\Bigl(1+\frac{4n^2r_+^2}{(r_+^2+a^2-n^2)^2}\Bigr)}{4\pi r_+(r_+^2+n^2+a^2)}\,, \label{Tel}\\
J&=J_0\Bigl(1+\frac{Q_e \phi_e}{r_+}\Bigr)\,,\quad J_0=\frac{a(r_+^2+n^2+a^2)}{2r_+}\,,\quad \Omega_H=\frac{a}{r_+^2+a^2+n^2}\,,\\
\psi&=\frac{1}{8\pi n}\,,\quad N=-\frac{4\pi n^3}{r_+}\Bigl(1+\frac{e^2(a^2-3r_+^2-n^2)}{(r_+^2+a^2-n^2)^2}\Bigr)\,,\label{Nel}
\end{align}
where $J$ is given by the horizon angular momentum $J_0$ \cite{Bordo:2019rhu, Clement:2019ghi} upgraded to the presence of electromagnetic field.

The {\em magnetic law} is obtained by setting the horizon electric charge in the second law above,
\be\label{1a}
Q_e^H=0\quad \Rightarrow \quad e=\frac{4gn^2r_+}{r_+^2+a^2-n^2}\,.
\ee
Upon doing this, the magnetic piece in \eqref{G1} vanishes and the following magnetic-type relations hold:
\begin{align}
\delta M&=T\delta S+\Omega_H \delta \tilde J+\phi_m \delta Q_m+\psi\delta \tilde N\,,\\
M&=2(TS+\psi \tilde N+\Omega_H \tilde J)+ \phi_m Q_m\,.
\end{align}
The thermodynamic quantities simplify significantly and we have
\begin{align}
\tilde G&=\frac{m}{2}+\frac{2n^2g^2r_+}{r_+^2+a^2-n^2}=\frac{1}{2}(m-\phi_m Q_m)\,,\\
\tilde M&=m=\frac{r_+^2+a^2-n^2}{2r_+}+\frac{2n^2g^2[(r_+^2+a^2+n^2)^2-4a^2n^2]}{r_+(r_+^2+a^2-n^2)^2}\,.\\
\phi_m&=-\frac{2gnr_+}{r_+^2+a^2-n^2}=-\frac{e}{2n}\,,\quad Q_m=-2ng\,,\\
S&=\pi (r_+^2+a^2+n^2)\,,\quad T=\frac{r_+^2+n^2-a^2-4n^2g^2\Bigl(1+\frac{4n^2r_+^2}{(r_+^2+a^2-n^2)^2}\Bigr)}{4\pi r_+(r_+^2+n^2+a^2)}\,,\\
\tilde J&=J_0\Bigl(1+\frac{\phi_mQ_m }{r_+}\Bigr)\,,\quad \Omega_H=\frac{a}{r_+^2+a^2+n^2}\,,\\
\psi&=\frac{1}{8\pi n}\,,\quad \tilde N=-\frac{4\pi n^3}{r_+}\Bigl(1+\frac{4n^2g^2(a^2-3r_+^2-n^2)}{(r_+^2+a^2-n^2)^2}\Bigr)\,.
\end{align}
Note that (apart from $G$ and $\tilde G$) all these quantities are related to the electric ones \eqref{Gel}--\eqref{Nel} by the electromagnetic duality
\eqref{EMduality}.

\begin{figure}
\begin{center}
\includegraphics[width=0.65\textwidth,height=0.32\textheight]{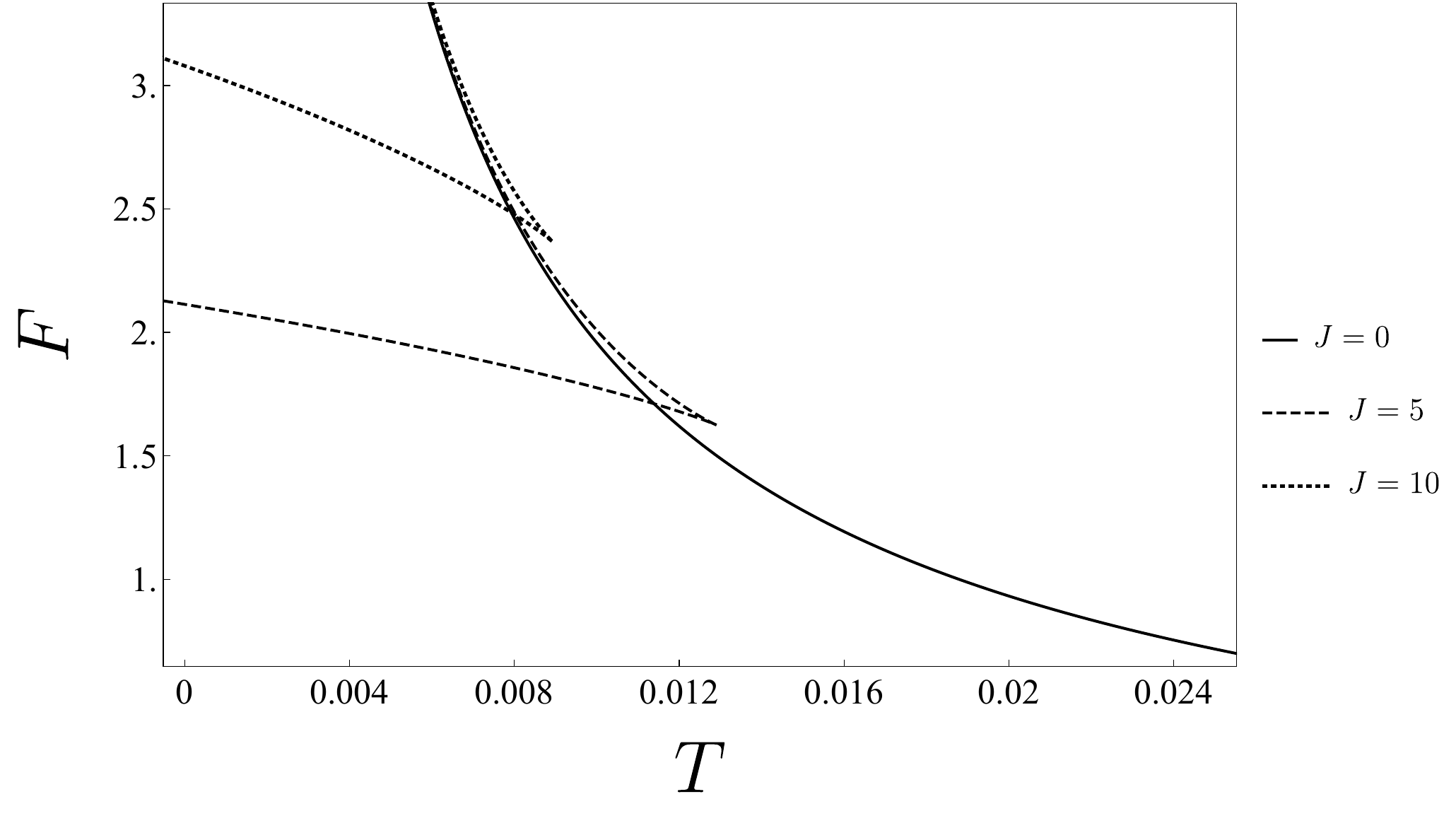}
\caption{{\bf Free energy: zero charge.} The $F-T$ diagram is displayed for fixed {$n=1$} and various angular momenta $J$.
We observe a very similar behavior to what happens for the Schwarzschild ($J=0$) black hole and the Kerr black hole $(J\neq 0$) respectively.
}\label{fig1}
\end{center}
\end{figure}

\begin{figure}
\begin{center}
\includegraphics[width=0.65\textwidth,height=0.32\textheight]{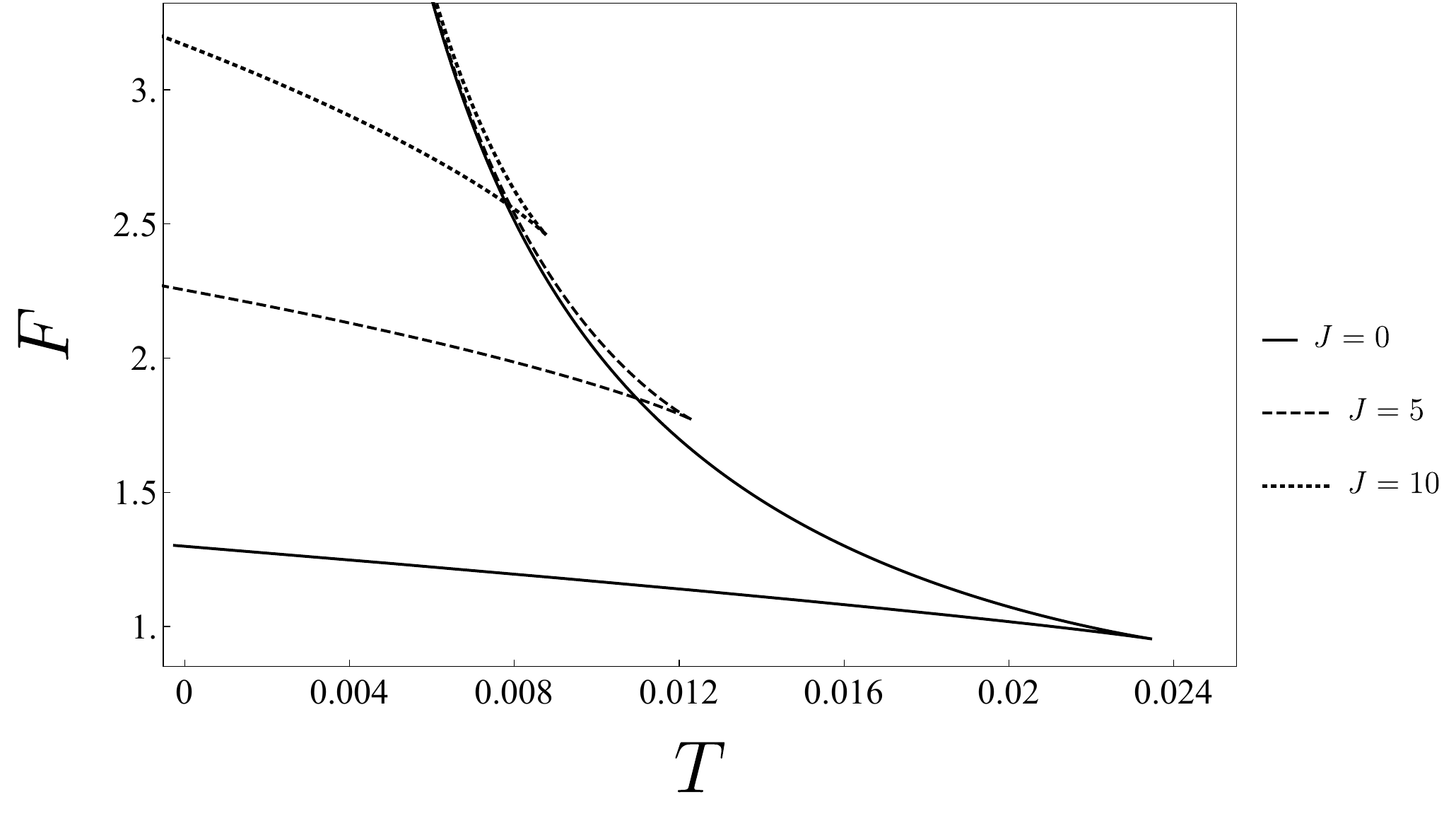}
\caption{{\bf Gibbs free energy: charged case.} The $F-T$ diagram is displayed for fixed {$n=1$}, $Q_e=1$, and various angular momenta.
The behavior of $F$ reminds that of the charged(--rotating) black hole.
}\label{fig2}
\end{center}
\end{figure}

To further probe the physical meaning of the obtained thermodynamic quantities, one can study the behavior of the corresponding free energy. For simplicity, we concentrate on the electric case and, similar to \cite{Ballon:2019uha}, work in an ensemble where we fix the Misner potential $\psi$ (and thence the NUT parameter $n$) instead of the Misner charge $N$ (which is quite algebraically complicated). Thus, starting from $G$ given by \eqref{Gel}, we consider the corresponding free energy
\ba
F&=&F(T, \psi, J, Q_e)=G+\phi_e Q_e+\Omega_H J\nonumber\\
&=&\frac{1}{4r_+}\Bigl(r_+^2+3a^2-n^2+\frac{e^2(3r_+^4+n^4+3a^4-4n^2a^2+6r_+^2a^2}{(r_+^2+a^2-n^2)^2}\Bigr)\,,
\ea
upon using \eqref{Gel}---\eqref{Nel}.
This is displayed as a function of temperature \eqref{Tel} in Fig.~\ref{fig1} for a purely rotating NUT with $Q_e=0$ and in  Fig.~\ref{fig2} for $Q_e=1$ and various
angular momenta.  It is encouraging to observe that in both cases the free energy has a similar behavior to what happens in the charged/rotating black hole case without a NUT parameter, e.g. \cite{Altamirano:2014tva}.

\section{Conclusions}

In this paper we have finalized  the work on thermodynamics of asymptotically flat Lorentzian Taub-NUT spacetimes started in \cite{Kubiznak:2019yiu}, by including simultaneously the effect of electromagnetic charges and black hole rotation.
Specifically, we have shown that one can formulate two first laws for the rotating NUTty dyon that involve either the horizon electric or the horizon magnetic charges in a fashion similar to \cite{Ballon:2019uha}. This was done by calculating the free energy using the on-shell Euclidean action. Then, armed with the knowledge of certain thermodynamic quantities and imposing the first law for the free energy, we solved for the Misner charges, angular momentum and conjugate magnetic potential. By constraining the value of the charges at the horizon, the law can be further reduced to obtain relatively simple purely electric and magnetic laws that are related by the electromagnetic duality.

As a result, we have fully characterized the thermodynamics of asymptotically flat Lorentzian Taub-NUT spacetimes. However, several questions are still left unanswered. One particular feature which deserves attention in the future is the appearance of horizon and asymptotic charges. This `breaks down' a status quo that the only horizon Noether charge that appears in the black hole thermodynamics first law is the entropy. Another puzzling feature to the authors is the fact that the magnetic potential that appears in the first law is not electromagnetically dual to the electric potential. Is this a consequence of the interplay between the Misner and Dirac strings in the presence of rotation? An obvious future direction would be to extend present results to more general spacetimes with NUT charges.
Particularly challenging seems the extension to the rotating AdS case (even in the absence of dyonic charges).  This would allow one to study the thermodynamic volume of the rotating NUTty solutions and probe the validity of the reverse isoperimetric ineqality conjecture \cite{Cvetic:2010jb}. Another direction would be to consider higher-dimensional Taub-NUT spacetimes in Einstein--Maxwell \cite{Mann:2005mb, Awad:2005ff, Dehghani:2006dk} or more general theories \cite{Dehghani:2005zm, Dehghani:2006aa, Dotti:2007az, Hendi:2008wq, Clarkson:2002uj, Bueno:2018uoy}.

\section*{Acknowledgements}
\label{sc:acknowledgements}
We would like to thank Robie A. Hennigar for discussions at the early stages of this project. The work was supported by the Perimeter Institute for Theoretical Physics and by the Natural Sciences and Engineering Research Council of Canada. Research at Perimeter Institute is supported in part by the Government of Canada through the Department of Innovation, Science and Economic Development Canada and by the Province of Ontario through the Ministry of Colleges and Universities.

\appendix

\section{Euclidean action calculation}\label{AppFE}

To find the Euclidean action for the asymptotically flat solution \eqref{eq:Met} studied in the main text, we will consider the rotating AdS NUTty dyon \cite{Griffiths:2005qp}, calculate its action by using the standard AdS counterterms, and take the asymptotically flat limit.

The rotating AdS NUTty dyonic solution reads
\ba\label{eq:Met2}
ds^2&=&-\frac{\Delta}{U} (e^1)^2+\frac{U}{\Delta}dr^2+\frac{U}{\left(1-y^2\right) P(y)}dy^2 +\frac{\left(1-y^2\right) P(y)}{U} (e^2)^2\;,\\
A&=&-\frac{e r+g(r^2+a^2-n^2)}{U}e^1+\frac{ag(1-y^2)}{U}e^2\;.
\ea
coordinate $y$ is related to the azimuthal angle $\theta$ by $y=\cos\theta$, $U=r^2+(n+ay)^2$,
\be
e^1=dt+\left(2 (s +ny)-a \left(1-y^2\right)\right)\frac{d\varphi}{K}\,,\quad
e^2=a dt-\left(r^2+n^2+a^2-2 a s \right)\frac{d\varphi}{K}\,,
\ee
and the metric functions $\Delta$ and $P(y)$ take a relatively complicated form:
\begin{align}
	\Delta&=a^2+e^2+4 g^2 n^2-n^2-2 m r+r^2-\frac{\Lambda}{3}\Bigl[r^2(a^2+6 n^2)+3n^2 (a^2-n^2)+r^4\Bigr]\;,\\
	P(y)&=1+\frac{\Lambda}{3}\left(a^2 y^2+4 a n y\right)\;.\label{A5}
\end{align}
The meaning of parameters $a, e, g, n, m, s$ is similar to the main text, while the new parameter $\Lambda=-3/\ell^2$ is the cosmological constant.

The coordinate $\phi$ is assumed $2\pi$ periodic; the role of $K$, which is not a new parameter but rather a specific function of the above parameters, is to ensure the absence of conical deficits on the north pole/south pole axes. In order to find its explicit form, one needs to probe the behavior of the angular part of the metric near the axes $y=\pm 1$, setting $r=$const. In order this is possible, one needs to eliminate the first (dominant) term in the metric, $-\frac{\Delta}{U} (e^1)^2$. It means that we have to set
\be\label{A6}
dt=-\left(2 (s +ny)-a \left(1-y^2\right)\right)\frac{d\varphi}{K}\,,
\ee
which is possible as both $t$ and $\varphi$ are timelike coordinates close to the axes.
Upon employing this condition, the angular part of the metric reads
\ba
ds^2&\approx& \frac{U}{\left(1-y^2\right) P(y)}dy^2 +\frac{\left(1-y^2\right) P(y)}{U} (e^2)^2\nonumber\\
&\approx&\frac{U}{(1-y^2)P}\Bigl[dy^2+\frac{(1-y^2)^2P^2}{K^2}d\varphi^2\Bigr]\,.
\ea
Requiring that $P^2/K^2=1$ on $y=\pm 1$, means that one needs to set
\be
K=K_\mp\,,\quad K_\pm=1\pm \frac{4na}{\ell^2}-\frac{a^2}{\ell^2}\,,
\ee
upon which the north/south axis is conical singularity free. Obviously, in the presence of $\Lambda$ this cannot be satisfied simultaneously and there will always be cosmic strings (conical deficits) present in the spacetime (in addition to Misner strings). However,
when the cosmological constant vanishes, $K$ is no longer needed and we can set
\be
K=1\,,
\ee
eliminating the cosmic strings from the spacetime.

The Euclidean action is calculated with the usual counter terms \cite{Emparan:1999pm}:
\begin{align}
I&=\frac{1}{16\pi}\int_{M}d^{4}x\sqrt{g}\left[ R+\frac{6}{\ell^{2}}-F^2\right]
+ \frac{1}{8\pi}\int_{\partial M}d^{3}x\sqrt{h}\left[\mathcal{K}
- \frac{2}{\ell} - \frac{\ell}{2}\mathcal{R}\left( h\right) \right]\,.
\label{action}
\end{align}
Here, $\ell^2=-3/\Lambda$, and $\mathcal{K}$ and $\mathcal{R}\left( h\right)$ are respectively the extrinsic curvature and Ricci scalar of the boundary. As always, in order to keep the metric and the vector potential real in the process of calculating the action one has to Wick rotate all of the following: the time coordinate $t$, the NUT parameter $n$, the rotation parameter $a$, and the electric $e$ and magnetic $g$ charge parameters. The action is associated with the corresponding free energy ${\cal G}=I/\beta$, $\beta=1/T$, where in the end the last four parameters $\{n,a,e,g\}$ are Wick rotated back, to yield
\be\label{Gfull}
\mathcal{G}=\frac{M}{2K}-\frac{\rh(a^2+3\rh^2+3n^2)}{2 K \ell^2}-\frac{\rh}{2}\frac{(\rh^2-n^2+a^2) \left(e^2-4 g^2 n^2\right)-8 e g n^2 \rh}{K\left[\rh^2+(a+n)^2\right]\left[\rh^2+(a-n)^2\right]}\,.
\ee
In particular, upon the limit $\Lambda\to 0$, or equivalently $\ell \to \infty$, and setting $K=1$, we recover the formula \eqref{G} in the main text.


\providecommand{\href}[2]{#2}\begingroup\raggedright\endgroup

\end{document}